\documentclass[conference]{IEEEtran}
\IEEEoverridecommandlockouts
\usepackage{cite}
\usepackage{amsmath,amssymb,amsfonts}
\usepackage{multirow}
\usepackage{algorithmic}
\usepackage{graphicx, graphics}
\usepackage{textcomp}
\usepackage{xcolor}
\usepackage{hyperref}

\def\BibTeX{{\rm B\kern-.05em{\sc i\kern-.025em b}\kern-.08em
    T\kern-.1667em\lower.7ex\hbox{E}\kern-.125emX}}
    
\makeatletter
\def\@IEEEpubidpullup{8\baselineskip}
\makeatother
\begin{document}
\IEEEoverridecommandlockouts
\IEEEpubid{
\parbox{\columnwidth}{\vspace{-6\baselineskip}Permission to make digital or hard copies of all or part of this work for personal or classroom use is granted without fee provided that copies are not made or distributed for profit or commercial advantage and that copies bear this notice and the full citation on the first page. Copyrights for components of this work owned by others than ACM must be honored. Abstracting with credit is permitted. To copy otherwise, or republish, to post on servers or to redistribute to lists, requires prior specific permission and/or a fee. Request permissions from \href{mailto:permissions@acm.org}{permissions@acm.org}.\hfill\vspace{0.8\baselineskip}\\
{
\small\textit{ASONAM '19}, August 27-30, 2019, Vancouver, Canada \\
\copyright\space 2019 Association for Computing Machinery. \\
ACM ISBN 978-1-4503-6868-1/19/08 \$15.00\\
\url{http://dx.doi.org/10.1145/3341161.3343685}
}
\hfill}
\hspace{0.9\columnsep}\makebox[\columnwidth]{\hfill}}
\IEEEpubidadjcol

\title{Two Decades of Network Science \\ \Large{as seen through the co-authorship network of network scientists}}

\author{\IEEEauthorblockN{Roland Molontay and Marcell Nagy}

\IEEEauthorblockA{MTA-BME Stochastics Research Group, Hungary
}
\IEEEauthorblockA{Department of Stochastics, Budapest University of Technology and Economics, Hungary}
\IEEEauthorblockA{Faculty of Informatics, University of Debrecen, Hungary}
\IEEEauthorblockA{Email: molontay@math.bme.hu, marcessz@math.bme.hu}}

\maketitle

\begin{abstract}
Complex networks have attracted a great deal of research interest in the last two decades since Watts~\&~Strogatz,  Barab\'asi~\&~Albert and Girvan~\&~Newman published their highly-cited seminal papers on small-world networks, on scale-free networks and on the community structure of complex networks, respectively. These fundamental papers initiated a new era of research establishing an interdisciplinary field called network science. Due to the multidisciplinary nature of the field, a diverse but not divided network science community has emerged in the past 20 years. This paper honors the contributions of network science by exploring the evolution of this community as seen through the growing co-authorship network of network scientists (here the notion refers to a scholar with at least one paper citing at least one of the three aforementioned milestone papers). After investigating various characteristics of 29,528 network science papers, we construct the co-authorship network of 52,406 network scientists and we analyze its topology and dynamics. We shed light on the collaboration patterns of the last 20 years of network science by investigating numerous structural properties of the co-authorship network and by using enhanced data visualization techniques. We also identify the most central authors, the largest communities, investigate the spatiotemporal changes, and compare the properties of the network to scientometric indicators.
\end{abstract}


\section{Introduction}
Complex networks have been studied extensively since they efficiently describe a wide range of systems, spanning many different disciplines, such as Biology (e.g. protein interaction networks), Information Technology (e.g., WWW, Internet), Social Sciences (e.g., collaboration, communication, economic, and political networks), etc.  Moreover, not only the networks originate from different domains, but the methodologies of network science as well, for instance, it heavily relies on the theories and methods of graph theory, statistical physics, computer science, statistics, and sociology.

In the last two decades, network science has become a new discipline of great importance. It can be regarded as a new academic field since 2005 when the U.S. National Research Council defined network science as a new field of basic research~\cite{national2005Network}. The most distinguished academic publishing companies announce the launch of new journals devoted to complex networks, one after another (e.g. Journal of Complex Networks by Oxford University Press, Network Science by Cambridge University Press, Applied Network Science and Social Network Analysis and Mining by Springer). Network science has its own prestigious conferences attended by thousands of scientists. Leading universities also continuously establish research centers and new departments for network science, furthermore, launch Master and PhD programs in this field (such as Yale University, Duke University, Northeastern University or Central European University).

\begin{figure}
    \centering
    \includegraphics[width=0.85\linewidth]{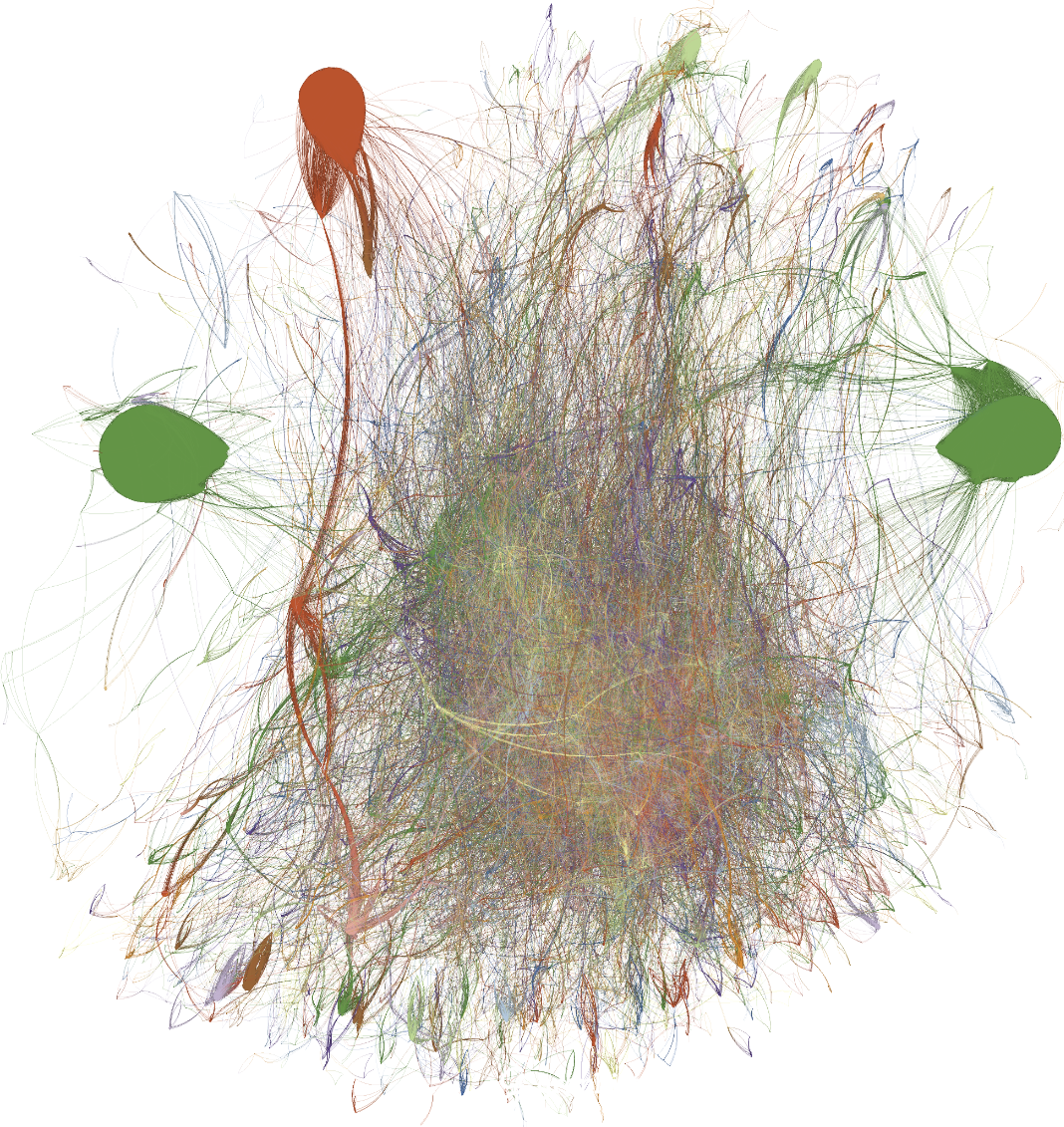}
    \caption{The largest connected component of the \textit{co-authorship network of network scientists} colored by communities.}
    \label{fig:NNS}
\end{figure}

The significance of network theory is also reflected in the large number of publications on complex networks and in the enormous number of citations of the pioneering papers by Barab\'asi~\&~Albert~\cite{barabasi1999emergence}, Watts~\&~Strogatz~\cite{watts1998collective} and Girvan~\&~Newman~\cite{girvan2002community} that turned the attention to complex networks. Some researchers interpret network science as a new paradigm shift~\cite{kocarev2010network}. However, complex networks are not only acknowledged by the research community, but innovative textbooks aimed for a wider audience have been also published~\cite{barabasi2016network,newman2018networks}, moreover, networks have appeared in the popular literature~\cite{barabasi2003linked,watts2004six} and mass media~\cite{connected2008} as well.

In the last two decades, complex networks became in the center of research interest thanks to --~among many others~-- the aforementioned three pioneering papers and due to the fact that the prompt evolution of information technology has opened up new approaches to the investigation of large networks. This period of 20 years can be regarded as the golden age of network science. The first challenge was to understand network topology, to this end, structural properties were put under the microscope one after the other (small-worldness, scale-free property, modularity, fractality, etc.) and various network models were proposed to understand and to mathematically describe the architecture and evolution of real-world networks~\cite{vespignani2018twenty}. In recent years, there has been a shift from the structural analysis to studying the control principles of complex networks~\cite{barabasi2019twenty}.  Remarkable computing power, massive datasets, and novel computational techniques keep great potential for network scientists for yet another 20 years~\cite{vespignani2018twenty}.

It is also important to mention some of the criticism leveled at network science. Some articles question the ubiquity of scale-free property and confute the assumption that biological networks or the internet are scale-free~\cite{broido2019scale,lima2009powerful,tanaka2005scale,willinger2009mathematics}. These papers claim that in a few works of network science, the data are insufficient and the measurements are not of satisfactory quality for the aim they are used for. Furthermore, these articles lack careful statistical tests~\cite{stumpf2012critical} and critics highlight that commonly used methods, such as least-squares fitting, can provide considerably inaccurate estimates of parameters~\cite{clauset2009power}. On the other hand, it is worth ascertaining that none of the critics doubt the importance of the study of complex networks, just raise concerns about certain methods and statements.

This work is a tribute to the achievements of the network science community in the past 20 years. We analyze 29,528 network science papers and  we also construct and investigate the co-authorship network of network scientists to identify how the network science community has been evolving over time. 

\section{Co-authorship network of network scientists}

In this study, we investigate the co-authorship network of network scientists, where a link between two scientists is formed by their co-authorship of at least one scientific paper. Co-authorship is one of the most important reflections of research collaboration, that is an essential mechanism that joins together distributed knowledge and expertise into novel discoveries. Co-authorship networks have been studied extensively in various ways and from various aspects, e.g. the collaboration network determined by the articles of a certain journal, a specific country or a research community that cites a particular influential paper~\cite{barabas2017impact,barabasi2002evolution,kumar2015co,newman2001structure,newman2004coauthorship}. To the best of our knowledge, co-authorship network of network scientists has been analyzed only by Newman \textit{et al.}~\cite{newman2006finding,newman2004finding}. However, their network consists of 1,589 authors, while this study investigates a much larger network (52,406 vertices and 329,181 edges) spanning a longer time horizon (1998-2019).  

We construct the co-authorship network of network scientists as follows. We consider three ground-breaking papers around the millennium that can be regarded as the roots of the rise of network science: the paper of Watts~\&~Strogatz~\cite{watts1998collective} on small-world networks, the work of Barab\'asi~\&~Albert~\cite{barabasi1999emergence} about scale-free networks and the paper of Girvan~\&~Newman~\cite{girvan2002community} that reveals the community structure of complex networks.

In this work, we consider a paper as a \textit{network science paper} if it cites at least one of the three aforementioned pivotal articles, similarly we call someone a \textit{network scientist} if (s)he has at least one \textit{network science paper}. The previous definitions of \textit{network science paper} and \textit{network scientist} are of course quite arbitrary. It is important to note that not all the papers that refer to one of the three seminal papers are necessarily about network science and there certainly exist network science articles that do not refer to any of the mentioned pioneering papers. On the other hand, we believe that this concept is a good proxy for our purposes and it is worth studying. We construct the co-authorship network of the \textit{network scientists} where two of them are connected if they have at least one joint \textit{network science paper} (see Fig.~\ref{fig:NNS}). The anonymized data of the constructed network is available in the supplementary material~\cite{supp}.

\begin{figure}[t]
    \centering
    \includegraphics[width=0.55\linewidth]{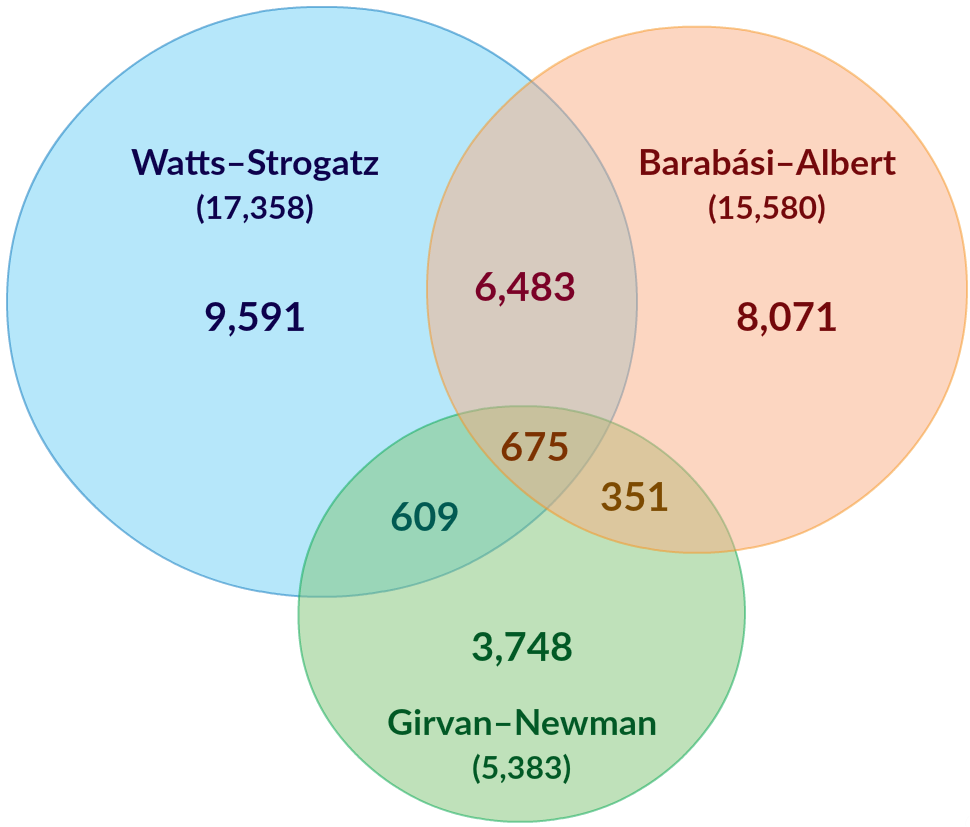}
    \vspace{-5pt}
    \caption{Distribution of the citations among the three pioneering papers.}
    \label{fig:venn}
\end{figure}

\section{Data collection and preparation}

We build our analysis on data collected from the Web of Science bibliographic database, retrieved on May 16, 2019. The collected data consist of 38,321 rows corresponding to the citing works of the three seminal articles~\cite{watts1998collective,barabasi1999emergence,girvan2002community}. For each citing paper we have information on the document title, publication name, publication type, publisher, publication year, authors full name, author address,  keywords, cited reference count, total times cited count, page count, abstract, etc.

After the data were collected, various data preparation steps were conducted, including merging the files, deleting duplicates and indicating which of the three seminal papers were cited by the given article. These preparation steps reduced the dataset to 29,528 unique rows and the citation pattern of the corresponding articles are shown in Fig.~\ref{fig:venn}.

The authors are represented by the full name field of Web of Science, however, this field is unfortunately not consistent, the author called John Michael Doe may appear as Doe,~John; John Doe; Doe, J.; Doe, J. M.; Doe, John Michael and other variants. To overcome this issue, we created a dictionary that defines the name variants that correspond to the same author. Furthermore, we cannot distinguish between different scientists with the same name, this issue is mainly relevant for Asian authors. However, the error introduced by this problem is negligible, as also pointed out by Newman~\cite{newman2001structure} and by Barab\'asi \textit{et al.}~\cite{barabasi2002evolution}.

\section{Analysis of network science papers}

\begin{figure}
    \centering
    \includegraphics[width=\linewidth]{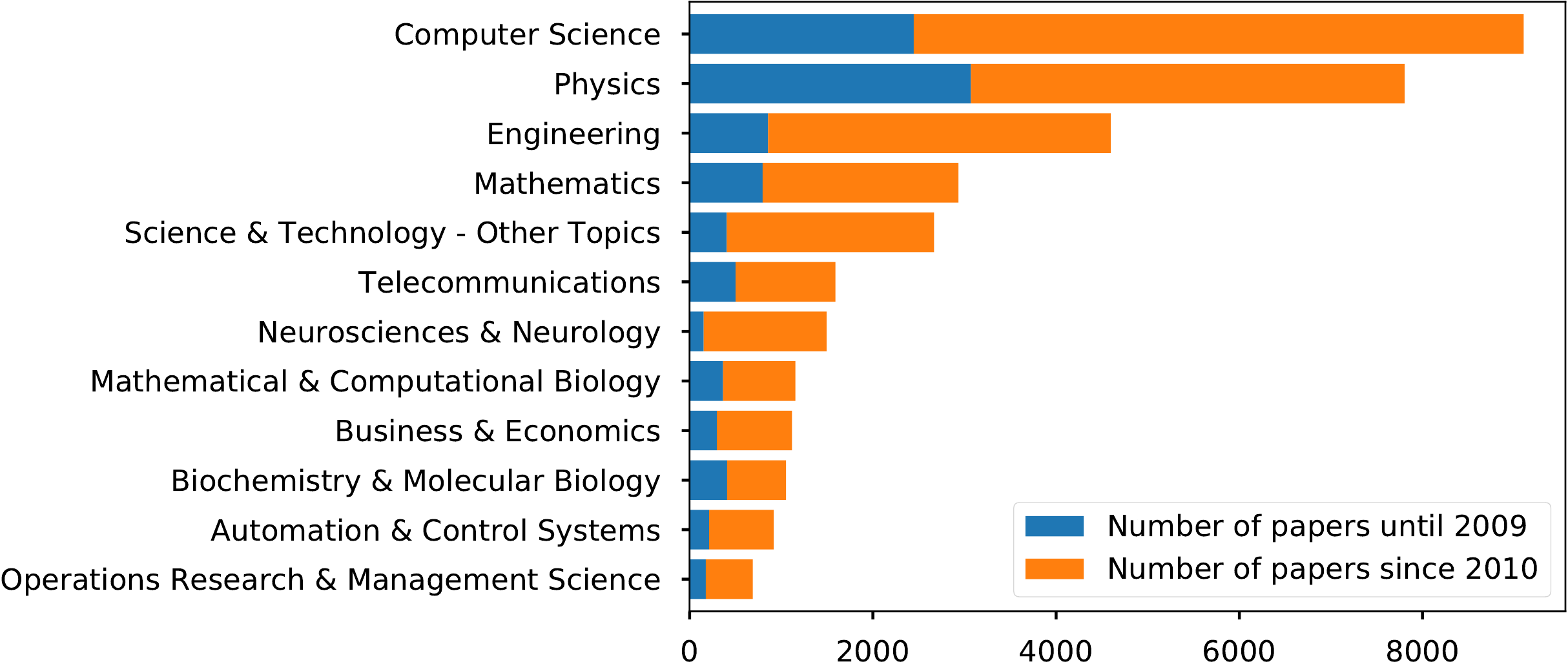}
    \vspace{-15pt}
    \caption{Top 12 research areas of \textit{network science papers} colored by the decade of publication time.}
    \label{fig:research areas}
\end{figure}

\begin{figure}
    \centering
    \includegraphics[width=\linewidth]{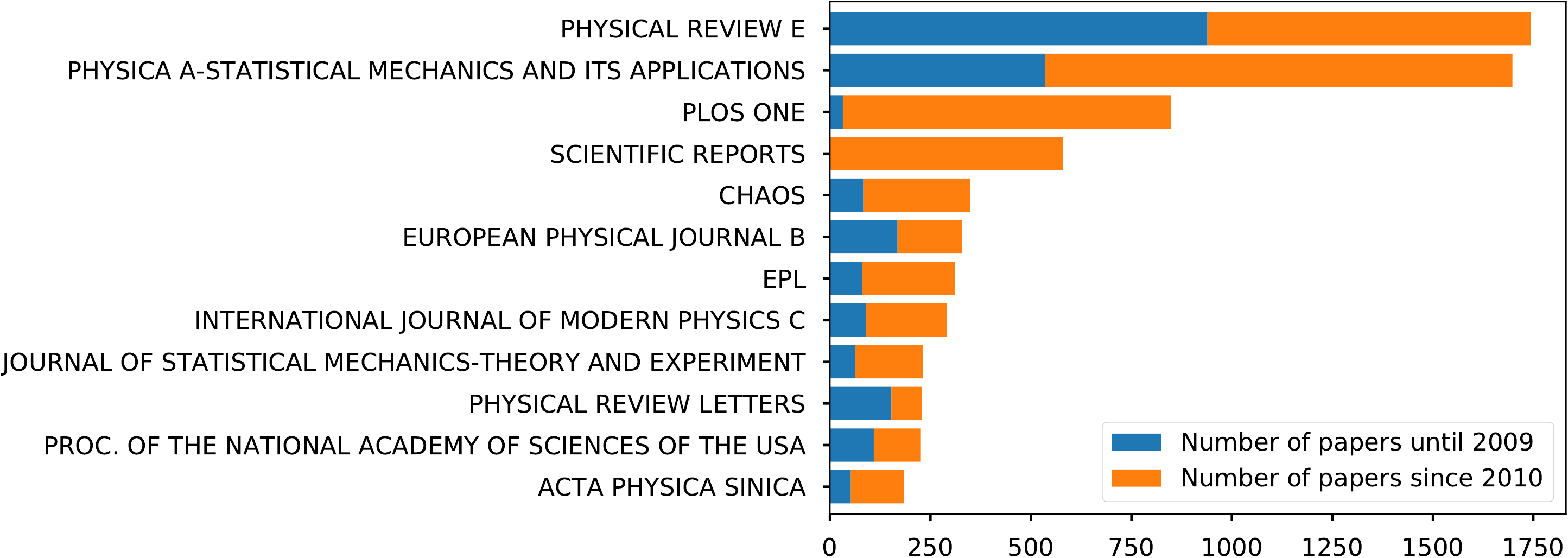}
    \vspace{-15pt}
    \caption{Top 12 journals of \textit{network science papers} colored by the decade of publication time.}
    \label{fig:journals}
\end{figure}

First, we analyze the enormous number of citing works, i.e. the \textit{network science papers}. Fig.~\ref{fig:research areas} shows the top 12 research areas that the citing works belong to, illustrating the interdisciplinary nature of network science. We can see that the first decade was dominated by physics while later computer science took over, it is also clear from the figure that neuroscience has started to use network science tools in the last decade. The journals that publish the most \textit{network science papers} are shown in Fig.~\ref{fig:journals}. Considering the number of publications, Physical Review E was the leading scientific forum of network science in the first half of the examined period, while Physica A, PLOS One and Scientific Reports have been catching up in the last decade.

\begin{figure}
    \centering
    \includegraphics[width=0.8\linewidth]{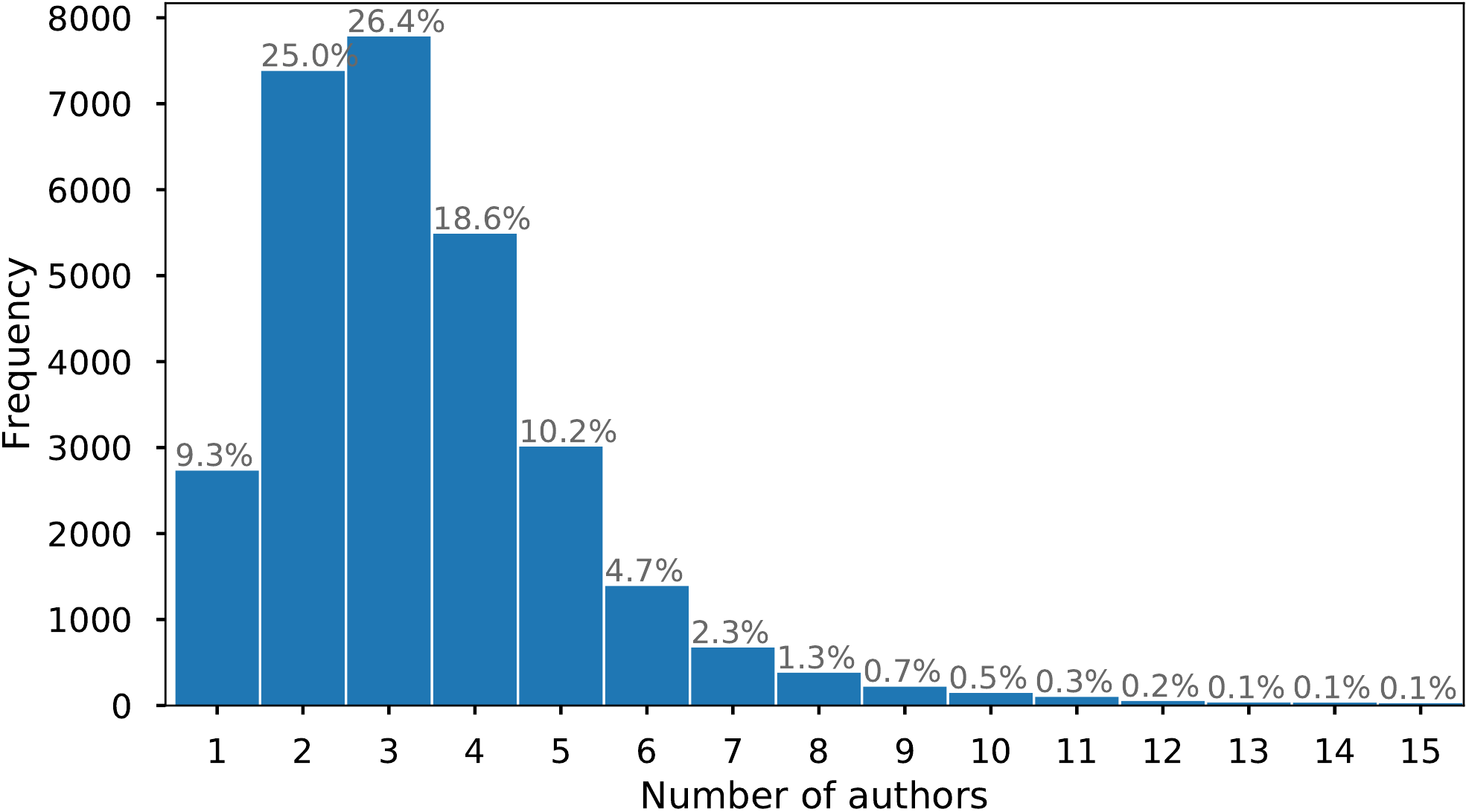}
    \caption{Histogram of the number of authors per paper (truncated at 15).}
    \label{fig:authors_per_paper}
\end{figure}

Fig.~\ref{fig:authors_per_paper} show the number of collaborating authors per citing works, the most typical numbers of co-authors in a \textit{network science paper} are 2 and 3. While the figure shows only up to 15 number of authors, there are a few papers with a high number of collaborating authors e.g. the paper with the highest number of authors (388) is a paper of the Alzheimer’s Disease Neuroimaging Initiative~\cite{lella2018communicability}. The authors of this article emerge as a maximal clique of the co-authorship network of \textit{network scientists} as it can be seen in Fig.~\ref{fig:NNS}. We also investigate the distribution of \textit{network science papers} written by \textit{network scientists}, the authors with the highest number of \textit{network science papers} together with the citation count of their \textit{network science papers} are shown in Table~\ref{top10}. The scientist with the highest number of papers is Guanrong Chen whose research areas are nonlinear systems and complex network dynamics and control.

\begin{table}[]
\vspace{-6pt}
\caption{Top 10 authors with the most network science papers.}
\label{top10}
\centering
\begin{tabular}{lcc}
\textbf{Name of author} & \multicolumn{1}{l}{\textbf{Number of papers}} & \multicolumn{1}{l}{\textbf{Number of citations}} \\ \hline
Guanrong Chen & 161 & 11,842\\
Tao Zhou & 132 &  8,548\\
Bing-Hong Wang & 125 &  4,739\\
Shlomo Havlin & 119 & 12,040\\
J\"urgen Kurths & 110 & 8,930\\
Eugene H. Stanley & 106 &  9,302\\
Zhongzhi Zhang & 99 &  1,935\\
Ying-Cheng Lai & 97 &  5,405\\
Albert-L\'aszl\'o Barab\'asi & 94 &  69,738\\
Matjaz Perc & 85 & 7,677
\end{tabular}
\end{table}

\begin{figure}
    \centering
    \includegraphics[width=0.9\linewidth]{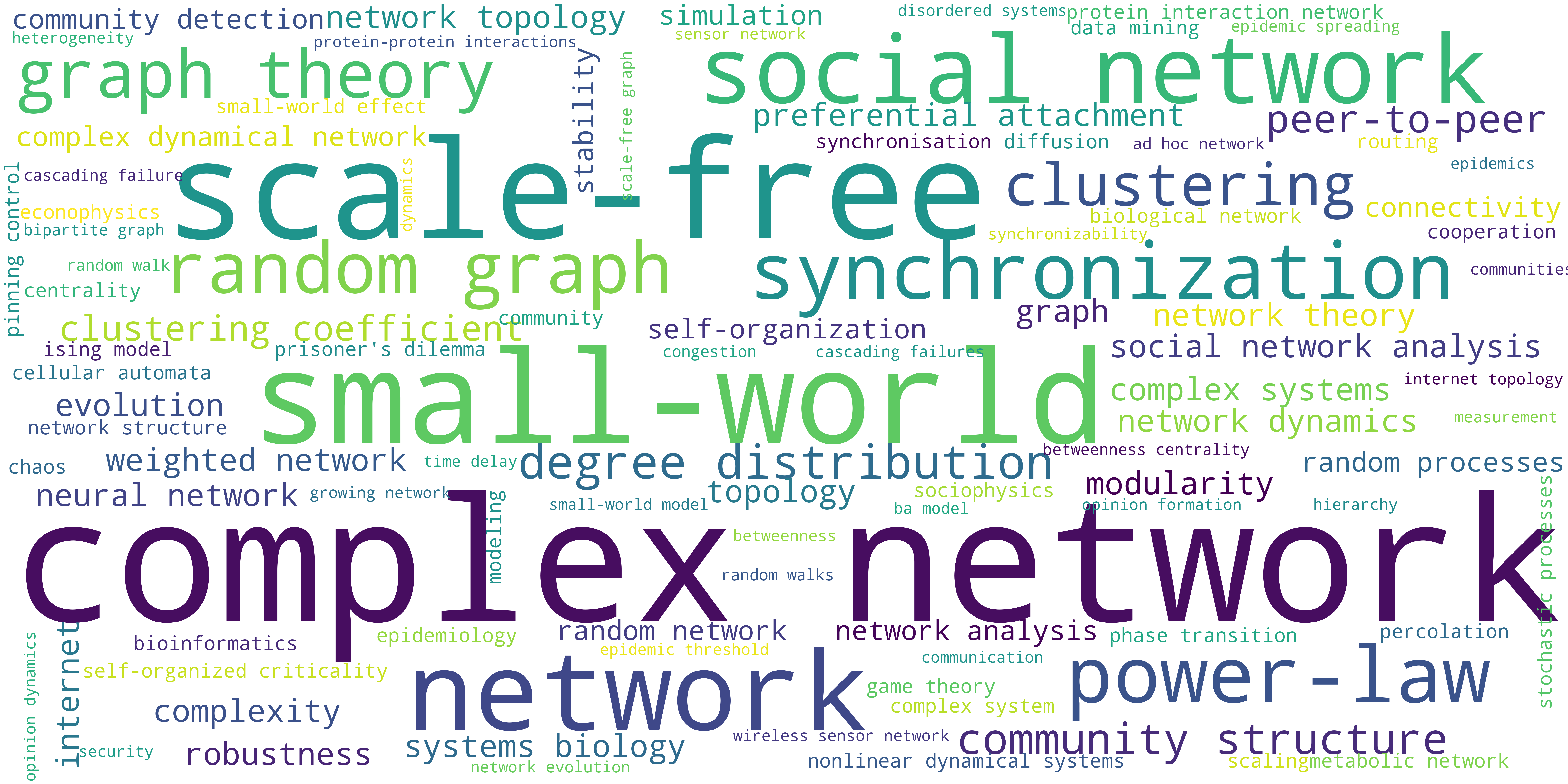}

    \vspace{0.4cm}

    \includegraphics[width=0.9\linewidth]{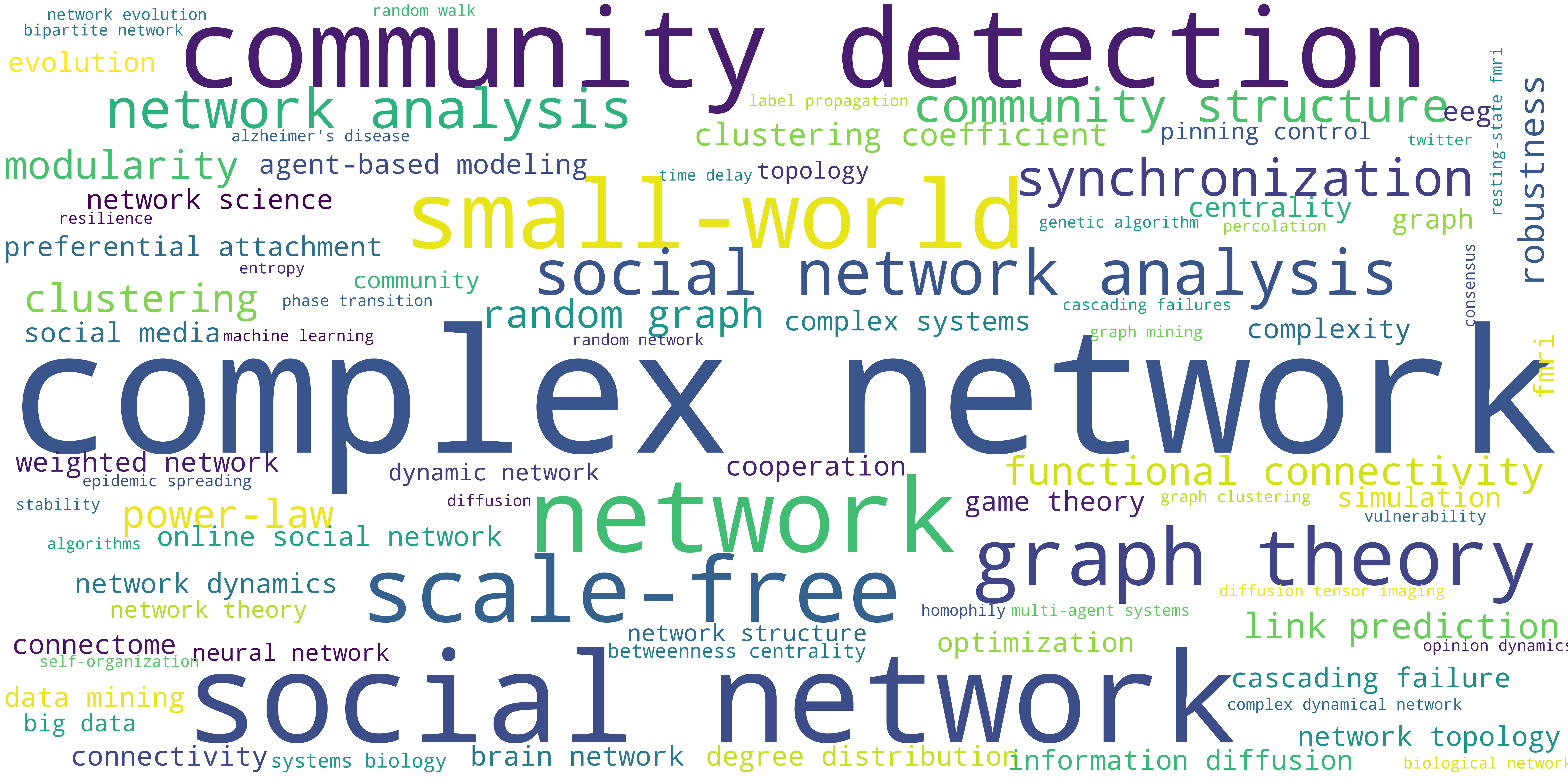}
    \vspace{-6pt}
    \caption{Word cloud of the most frequent keywords of \textit{network science papers} before 2009 (above) and since 2010 (bottom).}
    \label{wordclouds}
\end{figure}

Fig.~\ref{wordclouds} depicts separate world clouds of the most frequently used keywords of \textit{network science papers} written in the first and second decades of network science. We can observe that in the first half of the examined period structure related (e.g., scale-free, small world, topology) and modeling related keywords (e.g., preferential attachment, evolution, growing network, small-world model) dominated the study of complex networks, while in the last decade topics such as community detection, social network analysis, data-driven research (big data, data mining, link prediction, machine learning) have become more popular keywords. In the first period, the most studied real-world networks were Internet, peer-to-peer and protein interaction networks, however, since 2010 the research has tended to focus on online social networks and brain networks. Additional world clouds representing 5-year-long periods can be found in the supplementary material~\cite{supp}.

\begin{figure}
\vspace{-5pt}
    \centering
    \includegraphics[width=0.85\linewidth]{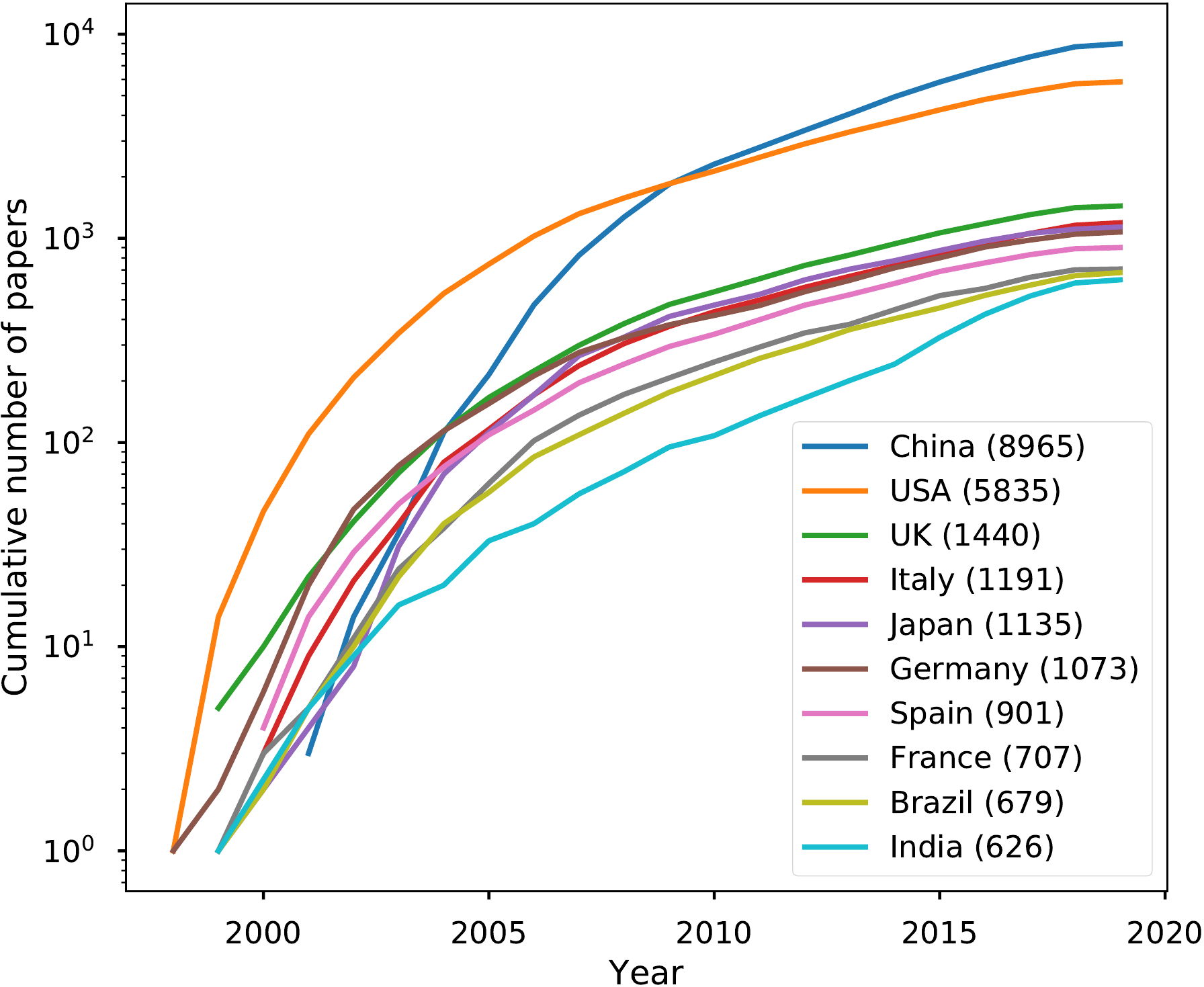}
    \vspace{-10pt}
    \caption{Cumulative number of \textit{network science papers} on a logarithmic scale by the country of the first author (only the Top 10 countries are shown).}
    \label{fig:cumsum}
\end{figure}

Based on the address of the first author, we identify the network science hot-spots and investigate the spatiotemporal changes. Fig.~\ref{fig:cumsum} demonstrates that China and USA are the two leading nations of network science with a fast increase of Chinese \textit{network science papers} in the last few years.

\section{Analysis of the co-authorship network}

The nodes of the co-authorship network of \textit{network scientists} correspond to the authors who have at least one \textit{network science papers} (i.e., a paper that cites at least one of the three seminal papers~\cite{barabasi1999emergence,watts1998collective,girvan2002community}), two of them are connected if they co-authored at least one \textit{network science paper}. The network is simple, undirected and unweighted meaning that here we do not measure the strength of the connection between two scientists by the number of their joint papers. The network has 52,406 nodes and 329,181 edges with an average degree of 12.56, however, the median degree is just 4. The largest connected component consists of 32,904 nodes and it is depicted in Fig. \ref{fig:NNS}.

\begin{figure}
    \centering
    \includegraphics[width=\linewidth]{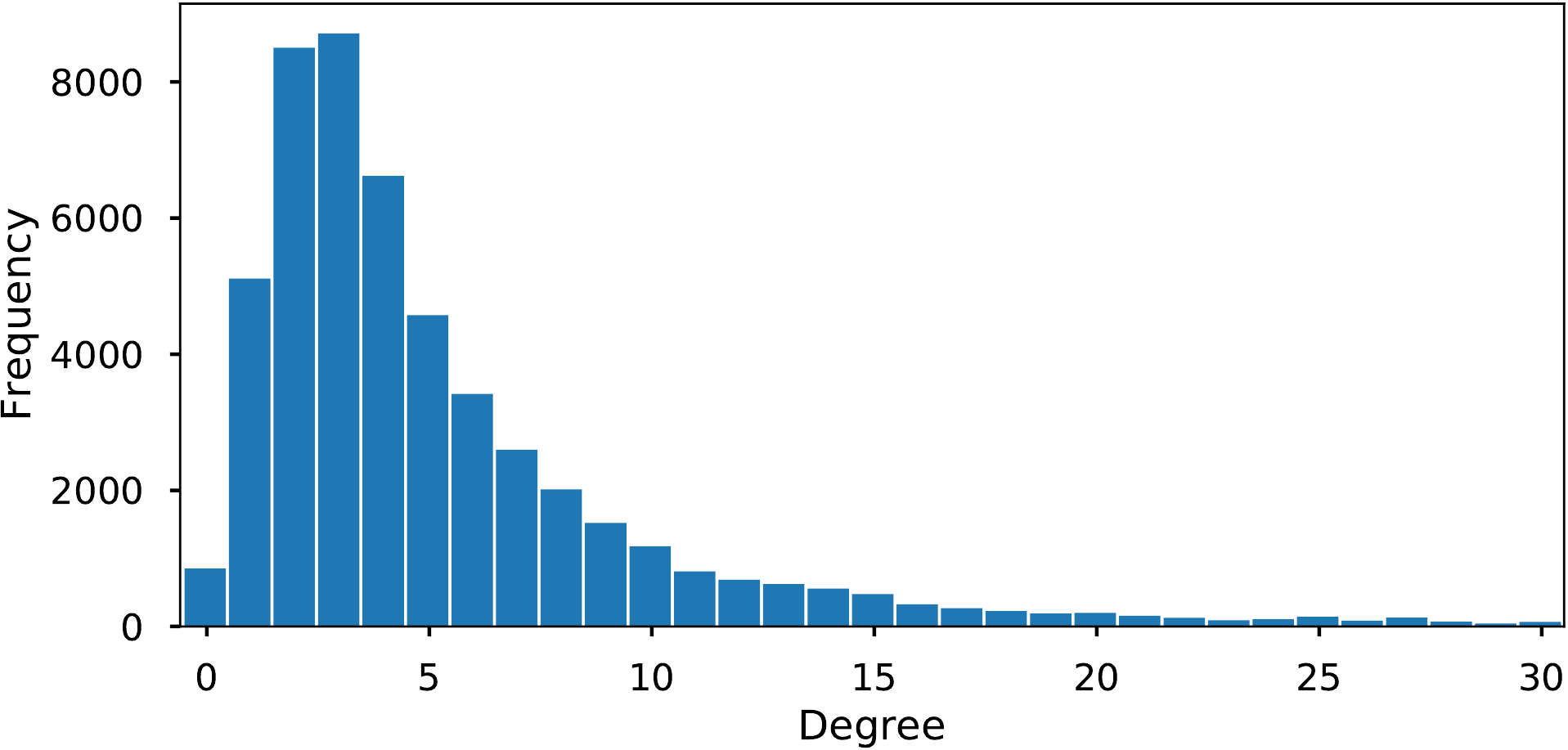}
    \vspace{-20pt}
    \caption{The degree distribution of the network (truncated at 30).}
    \label{fig:deg_dist}
\end{figure}

The degree distribution of the network is illustrated in Fig.~\ref{fig:deg_dist}. There are 851 isolated nodes in the graph (nodes with zero degrees), i.e. scholars who have a single-authored \textit{network science paper} but have not  co-authored any \textit{network science papers}. The most typical number of co-authors are between 2 and 4 and the tail of the distribution decays much slower than the number of authors per paper does (c.f. Fig.~\ref{fig:authors_per_paper}) since here the degree reflects all the number of co-authors who do not necessarily author the same paper. The highest degree is 444 corresponding to Paul M. Thompson neuroscientist, who is also an author of the paper with the highest number of co-authors~\cite{lella2018communicability}.

The network has a high assortativity coefficient of 0.57 that suggests that nodes tend to be connected to other nodes with similar degrees.  The co-authorship network is highly clustered with a global clustering coefficient of 0.98 and an average local clustering coefficient of 0.77. The fact that the average shortest path length in the largest connected component is 6.8 also supports the small-world nature of co-authorship networks.

\begin{table}[]
\caption{The top 10 authors with the highest centralities. Their ranks with respect to each metric are shown in brackets.}
\label{table:centralities}
\centering
\begin{tabular}{llll}
\multicolumn{1}{c}{\multirow{2}{*}{\textbf{Name}}} & \multicolumn{2}{c}{\textbf{Centralities}} & \multirow{2}{*}{\textbf{\begin{tabular}[c]{@{}c@{}}Number of\\ citations\end{tabular}}} \\
\multicolumn{1}{c}{} & \multicolumn{1}{l}{\textbf{Betweenness}} & \textbf{Harmonic} &  \\ \hline
J. Kurths & 0.027 (1) & 0.16 (1) & 8,930 (26) \\
E. H. Stanley & 0.024 (2) & 0.16 (2) & 9,302 (17) \\
G. Chen & 0.02 (3) & 0.16 (4) & 11,842  (14) \\
A.-L. Barab\'asi & 0.02 (4) & 0.15 (11) & 69,738 (1) \\
Y. He & 0.016 (5) & 0.15 (6) & 8,220 (296) \\
T. Zhou & 0.015 (6) & 0.16 (3) & 8,548 (294) \\
W. Wang & 0.013 (7) & 0.15 (8) & 328 (1840) \\
S. Havlin & 0.013 (8) & 0.15 (7) & 12,040 (12) \\
Z. Wang & 0.011 (9) & 0.15 (9) & 2,625 (389) \\
E. T. Bullmore & 0.011 (10) & 0.14 (47) & 15,486 (6)
\end{tabular}
\end{table}

\begin{figure}
\vspace{-6pt}
    \centering
    \includegraphics[width=0.9\linewidth]{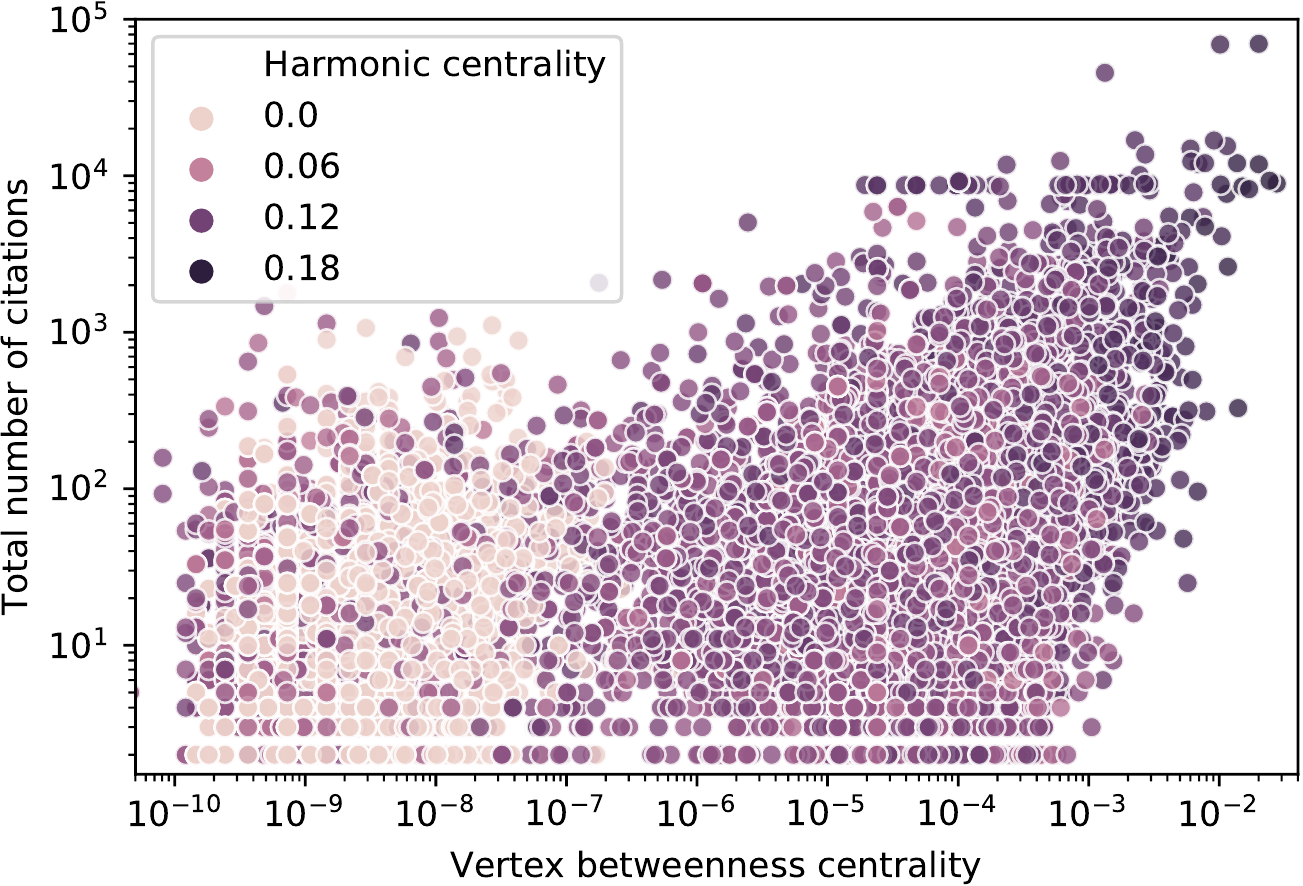}
    \vspace{-5pt}
    \caption{Relationship between centrality measurues of network scientists and the citation count of their network science papers.}
    \label{fig:centralities}
\end{figure}

To identify the most central authors of the network science community as seen through the co-authorship network, we calculate centrality measures such as betweenness and harmonic centralities of the nodes. The most central authors are shown in Table~\ref{table:centralities}. We also compare the centrality measures of the authors with the citation count of their \textit{network science papers}. Fig.~\ref{fig:centralities} shows the number of citations against the vertex betweenness centrality, colored by the harmonic centrality, we can conclude that there is a strong correlation between the authors' central role in the co-authorship network and their number of citations. 

\begin{figure}
    \centering
    \includegraphics[width=\linewidth]{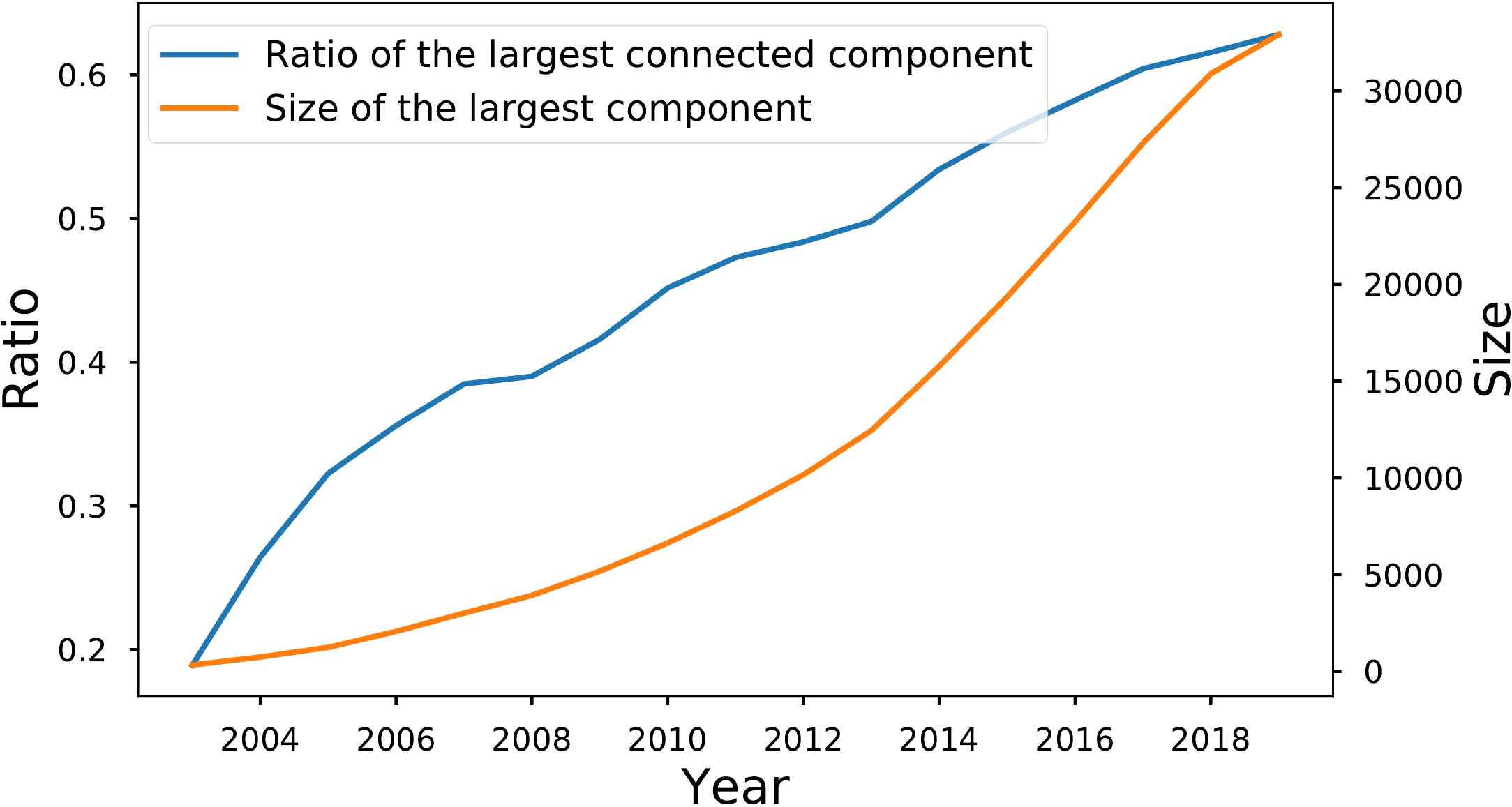}
    \vspace{-18pt}
    \caption{The absolute and relative size of the largest connected component of the co-authorship network.}
    \label{fig:size_and_comp}
\end{figure}

Network scientists have become more connected as time has gone by, as it is illustrated in Fig. \ref{fig:size_and_comp}, since not only the size of the largest component has increased over the years but also the ratio of the size of the giant component to the size of the entire network, indicating the emergence of a diverse but not divided network science community. The giant component consists of 32,904 nodes that are 62.8\% of the entire network size and it is illustrated in Fig.~\ref{fig:NNS}.

Using Clauset-Newman-Moore greedy modularity maximization community detection algorithm \cite{clauset2004finding}, we identify the dense subgraphs of the network. To retrieve some important discipline and location-related characteristics of the largest communities, we assigned a research area and a country for each author as the majority of the research areas corresponding to their papers and the most frequent country of their affiliations respectively.  The compositions of the ten largest communities are shown in Table~\ref{tab:comm}. The largest community consists of 14,136 authors dominated by Chinese physicists. We can observe that the smaller the communities are, the more homogeneous they are. For example, there is a community with 61\% European scientists and 43\% environmental scientists, moreover, the vast majority of the 10th largest community are  North American neuroscientists. 

\begin{table}[]
\caption{Composition of the largest communities.}
\label{tab:comm}
\scalebox{0.9}{
\begin{tabular}{rll}
\multicolumn{1}{l}{\textbf{Size}} & \textbf{Research area} & \textbf{Country} \\ \hline
14,136 & PHY 29\%, CS  25\%,  NN 9\% & \textbf{CHN 54\%}, EU 16\%, USA 12\% \\
1,394 & \textbf{CS 30\%}, PHY 13\%, NN 10\% & CHN 39\%,  USA 32\%, EU 11\% \\
1,302 & CS 21\%, BMB 15\%, PHY 8\% & EU 35\%, USA 26\%, JPN 9\% \\
848 & NN 31\%, CS 19\%, PHY 7\% & \textbf{USA 54\%}, EU 22\%, CHN 8\% \\
734 & \textbf{CS 39\%}, PHY 11\%, MAT 7\% & USA 28\%, EU 23\%, IND 13\% \\
520 & CS 23\%, BMB 14\%, PHY 12\% & \textbf{EU 57\%}, USA 16\%,  CAN 5\% \\
519 & \textbf{ESE 43\%},  CS 11\%,  LSB 8\% & \textbf{EU 61\%},  USA 13\%, BRA 9\% \\
449 & NN 31\%, PHY 24\%,   CS 10\% & EU 37\%, KOR 21\%, USA 11\% \\
416 & \textbf{GH 78\%}, NN 4\%, BMB 3\% & EU 43\%, USA 35\%, CHN 8\% \\
415 & \textbf{NN 93\%}, MAT 3\%, CS 2\% & \textbf{USA 89\%}, CAN 7\%, EU 3\% \\
\hline
\multicolumn{3}{l}{\scriptsize{\textbf{BMB}: Biochemistry \& Molecular Biology, \textbf{CS}: Computer Science}} \\
\multicolumn{3}{l}{\scriptsize{\textbf{PHY}: Phyics, \textbf{ESE}: Environmental Sciences \& Ecology, \textbf{GH}: Genetics \& Heredity}} \\
\multicolumn{3}{l}{\scriptsize{\textbf{LSB}: Life Sciences \& Biomedicine, \textbf{MAT}: Mathematics, \textbf{NN}: Neurosciences \& Neurology}} 
\end{tabular}}
\end{table}

\begin{figure}
    \centering
    \includegraphics[width=\linewidth]{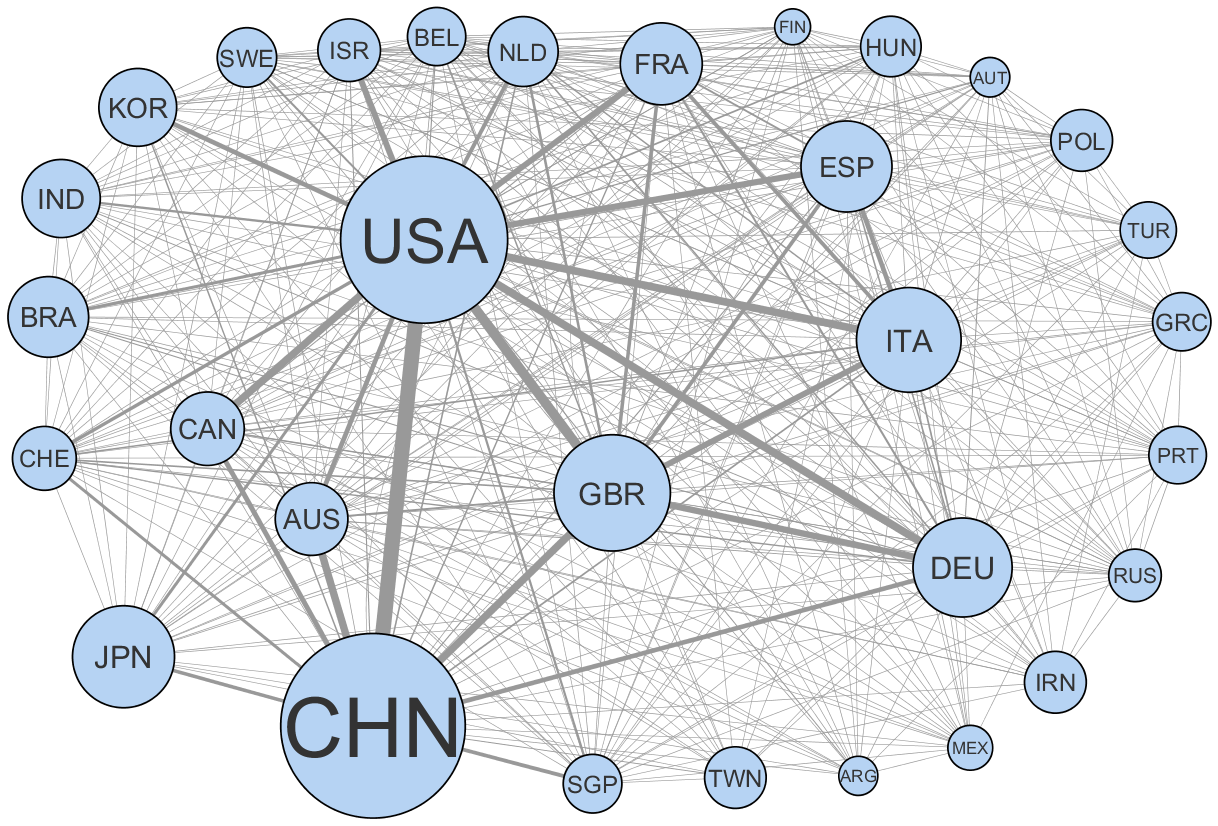}
    \vspace{-12pt}
    \caption{Network of international collaborations. The size of the node corresponds to the number of network science papers authored by at least one scientist from the corresponding country, the edge width indicates the number of papers written in the collaboration of authors from the corresponding countries. Only countries with at least 100 network science papers are shown in the figure.}
    \label{country_graph}
\end{figure}

\textit{Network scientists} come from 118 different countries which shows the international significance of network science. To illustrate the typical patterns of international collaborations, we created an edge-weighted network of countries with edge weights corresponding to the number of \textit{network science papers} that were written in the collaboration of at least one author from both countries (see Fig.~\ref{country_graph}). We can observe that while China has the highest number of \textit{network science papers} (see also Fig.~\ref{fig:cumsum}), US scientists wrote the most articles in international collaboration. It is also apparent that European countries collaborate with each other a lot.

\section{Conclusion}

Two decades ago a new multidisciplinary scientific field was born: network science. In this paper, we paid tribute to the network science community by investigating the past 20 years of complex network research as seen through the co-authorship network of network scientists. We studied 29,528 network science papers by extracting the distributions of research areas, journals, and keywords. Moreover, we constructed and extensively analyzed the co-authorship network of 52,406 network scientists, for example investigating its topological properties, namely its community structure, degree and centrality distributions. We also studied the spatiotemporal changes to provide insights on collaboration patterns. We compared the structural properties (e.g. centrality measures) with scientometric indicators (e.g. citation count) and found a high correlation.

After investigating the collaboration patterns and the increasing impact of complex networks, we are convinced that the next 20 years will produce at least as many fruitful scientific collaborations and outstanding discoveries in network science as the last two decades.

\section*{Acknowledgment}

We thank Bal\'azs Bir\'o and Ferenc P\'eter Kaiser for sharing their preliminary results with us.
The research reported in this paper was supported by the BME-Artificial Intelligence FIKP grant of EMMI (BME FIKP-MI/SC). The publication is also supported by the EFOP-3.6.2-16-2017-00015 project entitled "Deepening the activities of  HU-MATHS-IN, the Hungarian Service Network for Mathematics in Industry and Innovations". The research of R. Molontay was partially supported by NKFIH K123782 research grant.

\bibliographystyle{IEEEtran}
\bibliography{20ref.bib}
\end{document}